\documentclass[11pt]{article}

\usepackage{graphicx}

\newcommand{\rmt}[1]{\textrm{\tiny{#1}}}
\newcommand{\MK}{\ensuremath{\mathcal{K}} }
\newcommand{\MU}{\ensuremath{\mathcal{U}} }
\newcommand{\tMK}{\ensuremath{\tilde{\mathcal{K}}} }
\newcommand{\MSbar}{\ensuremath{\overline{{\rm MS}}}}
\def\nn{\nonumber}
\newcommand{\LMS}{\Lambda_{\rmt{\MSbar}}}

\begin{document}

\thispagestyle{empty}

\title{\vspace{1.3cm}{Comparison of NNLO and all-orders estimates of corrections to the GLS and Bjorken sum rules}}

\author{\textbf{Paul M. Brooks\footnote{e-mail:\texttt{brooks.pm@googlemail.com}}\,\, and C.J.
Maxwell\footnote{e-mail:\texttt{\texttt{c.j.maxwell@durham.ac.uk}}}}}

\date{\vskip10mm\emph{Institute for Particle Physics Phenomenology, University of Durham\\South Road, Durham, DH1 3LE,
England}}
\vspace{5cm}

\maketitle

\large\vspace{-8.45cm}

\hfill\vbox{\hbox{IPPP/06/52}\hbox{DCPT/06/104}
\hbox{August 2006}}

\normalsize\vspace{8.45cm}

\begin{abstract}
For the GLS and Bjorken DIS sum rules we compare fixed-order NNLO
perturbative QCD estimates, and all-orders resummed estimates, with
the available data, in order to assess the reliability of
fixed-order perturbation theory at rather small $Q^2$ values. Fits
are also performed for non-perturbative power corrections using a
recently proposed model.
\end{abstract}

\newpage
\setlength{\baselineskip}{18pt} \setlength{\parskip}{11pt}

\section{Introduction}\vspace{-11pt}
The perturbative series for QCD observables is not convergent, the
${n}^{\rm{th}}$-order coefficients exhibiting $n!$ growth at large $n$. In
the Borel plane there are singularities along the real axis: ultraviolet (UV)
renormalons along the negative real semi-axis, and infrared (IR) renormalons
lying on the integration contour, along the positive real semi-axis. The
latter render the Borel integral ambiguous, the ambiguity being structurally
identical to ambiguities arising in the non-perturbative operator product
expansion (OPE) \cite{r1}. Consequently,  all-orders perturbation theory is only
well-defined if supplemented by the non-perturbative OPE, allowing the
ambiguities to cancel.

In practice all-orders calculations are only possible in the
``leading-$b$ approximation'', in which  the ${n}^{\rm{th}}$-order
perturbative coefficient is recast as an expansion in powers of
$b=(33-2{N}_{f})/6$, the first beta-function coefficient (in SU$(3)$
QCD, with $N_f$ active quark flavours). The ``leading-$b$'' term
(proportional to $b^n$) is then used to approximate the
${n}^{\rm{th}}$-order coefficient \cite{r2,r3,r4}. Crucially this
term can be identified to all-orders from large $N_f$ calculations,
involving a restricted set of diagrams in which a chain of $n$
fermion bubbles is inserted into a basic ``skeleton'' diagram, in
all possible ways. In this way one can perform an all-orders
leading-$b$ resummation by explicitly evaluating the Borel
transform, and defining the Borel sum, suitably regulated to control
the IR renormalon ambiguities. Such results have been used in the
past to try to assess the likely accuracy of fixed-order
perturbation theory for quantities such as the
${R}_{{e}^{+}{e}^{-}}$ ratio at low values of the c.m.\@ energy
$Q^2$ (where the QCD coupling is not so small), and for its
tau-decay analogue $R_\tau$ (where again, at the scale of the $\tau$
mass, the coupling is relatively large) \cite{r2,r5,r5a,r6a,r6A}.
DIS sum rules have also been studied in this context
\cite{r6b,r6c,r6d,r6e,r6f,r6g,r6h}.

In matching to the exact fixed-order perturbative calculations (NLO and NNLO
coefficients are available for $R_{{e}^{+}{e}^{-}}$, $R_\tau$ and DIS sum rules), one faces
the problem that the all-orders leading-$b$ result is only
renormalization scheme (RS) invariant if the one-loop form of the coupling is
used, whereas the exact result must involve the higher-loop coupling. The matched
leading-$b$ resummed result consequently depends on the assumed
renormalization scale at which the matching takes place. To avoid this
ambiguity it has been argued in the past that one should remove scale
dependence by a resummation of scale logs to all-orders; employing the
effective charge approach \cite{r6}, or the ``complete renormalization
group improvement'' (CORGI) approach \cite{r7}. These approaches involve
scheme-invariants which can be approximated at the leading-$b$ level and
resummed \cite{r6a,r7a}.

In this paper we will use the CORGI approach to perform leading-$b$
resummations for DIS sum rules. The aim will be to assess the reliability of
fixed-order NNLO perturbation theory for these quantities, at the rather
small values of $Q^2$ for which they have been measured
\cite{r8,r9,r10,r11,r11a,r11b}. In addition we shall used a recently proposed model for
power corrections, based on the QCD skeleton expansion \cite{r12}, to obtain
fits for the non-perturbative contributions.

\section{Fixed-order and all-orders predictions}\vspace{-11pt}
The three physical quantities we are interested in are the GLS sum
rule \cite{r13}, the polarized Bjorken sum rule \cite{r14}, and the
unpolarized Bjorken sum rule \cite{r15}. These quantities can be
written as a parton model result plus perturbative corrections
(denoted by calligraphic letters),
\begin{eqnarray}
K_{GLS}(Q^2)&\equiv&\frac{1}{2}\int_{0}^{1}F_{3}^{\bar{\nu}p+\nu p}
(x,Q^{2})dx\nn\\
&=&3\Bigg{(}1-\frac{3}{4}C_{F}\MK(a)\Bigg{)}
\label{Kg},
\end{eqnarray}
\begin{eqnarray}
K_{pBj}(Q^2)&\equiv&\int_{0}^{1}g_{1}^{ep-en}(x,Q^{2})dx\nn\\
&=&\frac{1}{6}\Bigg{|}\frac{g_{A}}{g_{V}}\Bigg{|}\Bigg{(}1-\frac{3}{4}C_{F}\MK(a)\Bigg{)},
\label{Kp}
\end{eqnarray}
\begin{eqnarray}
U_{uBj}(Q^2)&\equiv&\int_{0}^{1}F_{1}^{\bar{\nu}p-\nu p}(x,Q^{2})dx\nn\\
&=&\Bigg{(}1-\frac{1}{2}C_{F}\MU(a)\Bigg{)}\label{Ub}.
\end{eqnarray}
where $|g_{A}/g_{V}|=1.257\pm0.003$ \cite{r15a}, and in the case of
$K_{GLS}(Q^{2})$ we have neglected contributions from
`light-by-light' scattering diagrams. The perturbative corrections
take the form of an expansion in powers of the strong coupling
$a=\alpha_{s}/\pi$,
\begin{eqnarray}
\MK(a)=a(Q^2)+\sum_{n=1}^{\infty}k_{n}a^{n+1}(Q^2),\qquad\qquad\qquad\MU(a)=a(Q^2)+\sum_{n=1}^{\infty}u_{n}a^{n+1}(Q^2).
\label{alo}
\end{eqnarray}
These expressions can be evaluated at
fixed-order, and the coefficients $k_{n}$ and $u_{n}$ are known up to $n=2$ \cite{r15b,r15c}
(NNLO).
\begin{eqnarray}
\MK_{\rmt{NNLO}}(a)=a+k_{1}a^{2}+k_{2}a^{3},\qquad\qquad\qquad\MU_{\rmt{NNLO}}(a)=a+u_{1}a^{2}+u_{2}a^{3}.\label{fxo}
\end{eqnarray}

\subsubsection*{The leading-$b$ approximation}\vspace{-11pt}
The leading-$b$ approximation arises from rearranging the expansion
as one in powers of $N_f$. We can write the ${n}^{\rm{th}}$-order
coefficients in the form,
\begin{equation}
{k}_{n}={k}_{n}^{[n]}{N}_{f}^{n}+{k}_{n}^{[n-1]}{N}_{f}^{n-1}+\ldots+{k}_{n}^{[0]}\;,
\end{equation}
and similarly for $u_n$. The leading $N_f$ coefficient ${k}_{n}^{[n]}$ can be computed exactly to all-orders
as noted above. One can replace $N_f$ by $(33/2-3b)$ to obtain an expansion in powers of $b$,
\begin{equation}
{k}_{n}={k}_{n}^{(n)}{b}^{n}+{k}_{n}^{(n-1)}{b}^{n-1}+\ldots+{k}_{n}^{(0)}\;.
\label{eq:leadingb}
\end{equation}
The leading-$b$ term ${k}_{n}^{(L)}\equiv {k}_{n}^{(n)}b^n$ can then
be used to approximate $k_n$ \cite{r2}. To see how well this
approximation works we can compare the known exact results for
$k_1$, $k_2$, and $u_1$, $u_2$, with the leading-$b$ approximations
${k}_{1}^{(L)}$, ${k}_{2}^{(L)}$ and ${u}_{1}^{(L)}$,
${u}_{2}^{(L)}$. We shall assume the \MSbar~scheme with
renormalization scale $\mu^2=Q^2$. For $\MK(a)$ one has the exact
coefficients
\begin{eqnarray}
k_1&=& -0.333 N_f+ 4.58
\\
k_2&=& 0.177{N}_{f}^{2}-7.61{N}_{f}+41.4.
\end{eqnarray}
To be compared with the leading-$b$ approximations
\begin{eqnarray}
{k}_{1}^{(L)}&=&-0.333{N}_{f}+5.5
\\
{k}_{2}^{(L)}&=&0.177{N}_{f}^{2}-5.86{N}_{f}+48.3.
\end{eqnarray}
The leading-$N_f$ coefficients of course agree exactly by
construction, but one sees agreement in sign and $\sim 20-30\%$
level agreement in magnitude for the sub-leading coefficients as
well. The corresponding coefficients for $\MU(a)$ are
\begin{eqnarray}
{u}_{1}&=&-0.444{N}_{f}+5.75
\\
u_{2}&=&0.239{N}_{f}^{2}-9.5{N}_{f}+54.2.
\end{eqnarray}
To be compared with the leading-$b$ approximations
\begin{eqnarray}
{u}_{1}^{(L)}&=&-0.444{N}_{f}+7.33
\\
{u}_{2}^{(L)}&=&0.239{N}_{f}^{2}-7.89{N}_{f}+65.1.
\end{eqnarray}
Again one sees that sub-leading coefficients are well-reproduced in
sign and magnitude. This can be partially understood since it is
possible to derive a fully normalized asymptotic result for the
${k}^{[n-r]}$ and ${u}^{[n-r]}$ coefficients for {\it large} $n$ and
{\it fixed} $r$. Such a result for the  QCD Adler function of vacuum
polarization, $D(Q^2)$, was derived in Ref.~\cite{r15d} from a
consideration of the four-fermion operators controlling ultraviolet
renormalon singularities. It can be easily extended to the DIS sum
rules since the same dominant four-fermion operator is involved
\cite{r15e}. If one writes the leading-$b$ term as
\begin{eqnarray}
{k}_{n}^{(L)}&=&{\hat{k}}_{n}^{[n]}{N}_{f}^{n}+{\hat{k}}_{n-1}^{[n-1]}{N}_{f}^{n-1}+\ldots+{\hat{k}}_{n}^{[0]}
\\
{u}_{n}^{(L)}&=&{\hat{u}}_{n}^{[n]}{N}_{f}^{n}+{\hat{u}}_{n-1}^{[n-1]}{N}_{f}^{n-1}+\ldots+{\hat{u}}_{n}^{[0]},
\end{eqnarray}
then one has the asymptotic result for large $n$ and fixed $r$
\begin{eqnarray}
{k}_{n}^{[n-r]}&\approx&{\hat{k}}_{n}^{[n-r]}\left(1+O\left(\frac{1}{n}\right)\right)+O\left(\frac{1}{N^2}\right)
\\
{u}_{n}^{[n-r]}&\approx&{\hat{u}}_{n}^{[n-r]}\left(1+O\left(\frac{1}{n}\right)\right)+O\left(\frac{1}{N^2}\right).
\end{eqnarray}
The $O(1/N^2)$ are terms sub-leading in the number of colours, $N$.
The asymptotics agree up to $O(1/n)$ terms in the ``planar
approximation'' \cite{r15d}, where at each order in $N_f$ only the
highest power of $N$ is retained. Of course this asymptotic result
does not explain the $\sim 20-30\%$ level agreement observed on
comparing Eqs.~(8), (9), (12) and (13) with Eqs.~(10), (11), (14)
and (15), for such small values of $n=1,2$. The agreement is also
remarkably good for the ${k}_{n}^{[0]}$ and ${u}_{n}^{[0]}$
coefficients which correspond to $N_f=0$, or the large-$N$ limit.
Similar remarks apply for the Adler function $D(Q^2)$ and some
speculative ideas as to how this might arise are given in
Ref.~\cite{r15d}. Whilst this good agreement for given powers in the
$N_f$ expansion is interesting, of more relevance is how well the
overall $k_1$, $k_2$ and $u_1$, $u_2$ coefficients are approximated,
since one would not expect the $\sim 20-30\%$ accuracy to survive
the addition of the terms. We give below the exact and leading-$b$
coefficients for $N_f=\{ 0,1,2,3,4,5\}$. For $\MK(a)$ one has
\begin{eqnarray}
{k}_{1}&=&\{ 4.58,4.25,3.92,3.58,3.25,2.92\}
\\
{k}_{1}^{(L)}&=&\{ 5.5,5.17,4.83,4.5,4.17,3.83\}
\\
{k}_{2}&=&\{ 41.4,34,26.9,20.2,13.9,7.84\}.
\\
{k}_{2}^{(L)}&=&\{ 48.3,42.6,37.3,32.3,27.7,23.5\}.
\end{eqnarray}
Whilst for $\MU(a)$ one has
\begin{eqnarray}
u_1&=&\{ 5.75,5.31,4.86,4.42,3.97,3.53\}
\\
{u}_{1}^{(L)}&=&\{ 7.33,6.89,6.44,6,5.56,5.11\}
\\
u_2&=&\{ 54.2,45,36.2,27.9,20.1,12.7\}
\\
{u}_{2}^{(L)}&=&\{ 65.1,57.5,50.3,43.6,37.4,31.6\}.
\end{eqnarray}
As expected from the above discussion one finds good $\sim 20\%$ level agreement for $N_f=0$ which gradually worsens
to factor of $2$ level agreement at $N_f=5$.

\subsubsection*{All-orders predictions}\vspace{-11pt}
The resummation of leading-$b$ terms to all-orders, which we shall
denote as $\MK_{\infty}(a)$, can be accomplished using the Borel sum
method,
 \begin{eqnarray}
B[{\MK}^{(L)}](z)=\sum_{n=0}^{\infty}\frac{k_{n}^{(L)}}{n!}z^{n}&\qquad\Longrightarrow\qquad&\MK_{\infty}(a)\simeq\int_{0}^{\infty}dz
e^{-z/a}B[{\MK}^{(L)}](z).
\label{BT}
\end{eqnarray}
The Borel transforms of the three above quantities in the
leading-$b$ approximation can be calculated from the results in
Refs.~\cite{r16} and \cite{r17}. One finds,
\begin{eqnarray}
B[\MK^{(L)}](z)=
\frac{4/9}{\Big{(}1+\frac{z}{z_{1}}\Big{)}}-\frac{1/18}{\Big{(}1
+\frac{z}{z_{2}}\Big{)}}+\frac{8/9}{\Big{(}1-\frac{z}{z_{1}}\Big{)}}-\frac{5/18}{\Big{(}1-
\frac{z}{z_{2}}\Big{)}},
\end{eqnarray}
and,
\begin{eqnarray}
B[\MU^{(L)}](z)=\frac{1/6}{\Big{(}1+\frac{z}{z_{2}}\Big{)}}+\frac{4/3}{\Big{(}1-\frac{z}{z_{1}}\Big{)}}-\frac{1/2}{\Big{(}1-\frac{z}{z_{2}}\Big{)}}.
\end{eqnarray}
The Borel transforms have simple pole singularities on the positive
$z$-axis (IR renormalons), and on the negative $z$-axis (UV
renormalons); the positions are at $z=z_n\equiv{2n}/{b}$. One can
easily compute the PV regulated leading-$b$ Borel sums,
${\MK}_{\infty}(a)$ and ${\MU}_{\infty}(a)$ \cite{r2},
\begin{eqnarray}
{\MK}_{\infty}(a)&=&\frac{1}{9b} \Bigg{[}-8e^{{z_1}/a(Q^2)}\textrm{Ei}(-z_{1}/a(Q^2))+2e^{z_{2}/a(Q^2)}\textrm{Ei}(-z_{2}/a(Q^2))\nonumber\\
&&+\;16e^{-z_{1}/a(Q^2)}\textrm{Ei}(z_{1}/a(Q^2))-10e^{-z_{2}/a(Q^2)}\textrm{Ei}(z_{2}/a(Q^2))\Bigg{]},
\label{eq:Kinf}
\end{eqnarray}
and,
\begin{eqnarray}
{\MU}_{\infty}(a)&=&\frac{1}{3b}\Bigg{[}8e^{-z_{1}/a(Q^2)}\textrm{Ei}(z_{1}/a(Q^2))-6e^{-z_{2}/a(Q^2)}\textrm{Ei}(z_{2}/a(Q^2))
\nonumber\\
&-&2e^{z_{2}/a(Q^2)}\textrm{Ei}(-z_{2}/a(Q^2))\Bigg{]}.
\label{eq:Uinf}
\end{eqnarray}
Ei$(x)$ is the exponential integral function defined (for $x<0$, UV renormalons) as,
\begin{eqnarray}
\textrm{Ei}(x)\equiv-\int_{-x}^{\infty}dt\frac{e^{-t}}{t}\;,
\end{eqnarray}
and for $x>0$, by taking the PV of the integral (IR renormalons).
For these all-orders resummations to be RS (scale)-invariant we need to take $a(Q^2)$ as the
one-loop coupling,
\begin{equation}
a(Q^{2})=\frac{2}{b\ln(Q^{2}/\Lambda^{2})}.
\label{eq:olc}
\end{equation}

The all-orders results of Eqs.~(\ref{eq:Kinf}) and (\ref{eq:Uinf})
are derived in the so-called $V$-scheme (\MSbar~with renormalization
scale ${\mu}^{2}={e}^{-5/3}{Q}^{2}$), and hence $\Lambda$ in
Eq.~(\ref{eq:olc}) will refer to that in the $V$-scheme defined by,
\begin{eqnarray}
\Lambda_{V}={\rm e}^{5/6}\LMS. \label{eq:Lv}
\end{eqnarray}

\subsubsection*{CORGI perturbation theory}\vspace{-11pt}

As we noted above there is a problem if we wish to match these
leading-$b$ resummations with exact higher-order NLO and NNLO
calculations for $\MK(a)$ and $\MU(a)$, which involve the
higher-loop coupling. We shall avoid the matching ambiguity by
employing the CORGI approach \cite{r7}. The standard RS-dependent
coupling $a(\mu)$ satisfies the RG-equation,
\begin{equation}
\frac{da(\mu)}{d\ln(\mu)}=\beta(a)=-b a^2 (1 + c a + c_2 a^2 + c_3 a^3 + \cdots)\;.
\end{equation}
Here $b=(33-2{N}_{f})/6$ and $c=(153-19{N}_{f})/12b$ are universal and RS-invariant,
the higher coefficients, ${c}_{n\ge{2}}$, are RS-dependent and may be used,
together with dimensional transmutation parameter, $\Lambda$, to label the
scheme. The CORGI approach
consists of resumming to all-orders the RG-predictable terms available to a given fixed-order
of calculation.
The standard perturbative expansion is then replaced with the following expression,
\begin{eqnarray}
\tilde{\MK}(a_{0})=a_{0}+\sum_{n=2}^{\infty}X_{n}a_{0}^{n+1},
\label{eq:CORGIallo}
\end{eqnarray}
which removes the renormalization scale dependence of
Eq.~(\ref{alo}). We will use a tilde to denote CORGI-ized results. Initially we
shall deal with $\MK(a)$ but the results are easily generalized to $\MU(a)$.

The coefficients $X_{n}$ are renormalization scheme invariant
quantities, each of which can be derived from an ${\rm N}^{n}$LO
calculation of the coefficients of $\MK(a)$,  and of the beta
function equation coefficients, $c_{n}$. For $n=2$ and 3 they have
the form,
\begin{eqnarray}
X_{2}&=&k_{2}-k_{1}^{2}-ck_{1}+c_{2},
\label{eq:X2}\\
X_{3}&=&k_{3}-3k_{1}k_{2}+2k_{1}^{3}+\frac{ck_{1}^{2}}{2}-k_{1}c_{2}+\frac{1}{2}c_{3}.
\label{eq:X3}
\end{eqnarray}
The coupling $a_{0}$ used in Eq.~(\ref{eq:CORGIallo}) has the
form \cite{r19,r20},
\begin{eqnarray}
a_{0}(Q)=\frac{-1}{c\Big{(}1+W_{-1}\Big{[}-\frac{1}{e}\Big{(}\frac{Q^{2}}{\Lambda_{\MK}^{2}}\Big{)}^{-b/2c}\Big{]}\Big{)}},\label{eq:aok}
\label{eq:a}
\end{eqnarray}
and is the solution to the beta function equation in the 't Hooft scheme
\cite{r21} (a scheme where $c_{n\geq 2}=0$). Here $W$ is the Lambert $W$
function, defined implicitly by $W(z){\exp}(W(z))=z$. The ``$-1$'' subscript
on $W$ denotes the branch of the Lambert $W$ function required for asymptotic
freedom, the nomenclature being that of Ref.~\cite{r22}.

As a result of adopting the CORGI approach, we use a new scale
parameter $\Lambda_{\MK}$ which is RS invariant but dependent on the
observable under consideration. It is related to the \MSbar~scale
parameter by,
\begin{eqnarray}
\Lambda_{\MK}&=&\tilde{\Lambda}_{\rmt \MSbar}\exp(k_{1}^{\rmt\MSbar}/b)\nn\\
&=&\Lambda_{\rmt\MSbar}\left(\frac{2c}{b}\right)^{-c/b}\exp(k_{1}^{\rmt\MSbar}/b).\label{eq:Lk}
\end{eqnarray}
Here ${k}_{1}^{\rmt\MSbar}$ denotes the NLO perturbative coefficient
with scale choice $\mu=Q$. The exponential term in the above
equation has the effect of resumming a set of RG predictable terms
present in the full perturbative expansion, it also reproduces the
$\mathcal{O}(a^{2})$ term present in Eq.~(\ref{alo}) but absent from
Eq.~(\ref{eq:CORGIallo}). The difference between $\Lambda$ and
$\tilde{\Lambda}$ is due to different conventions for integrating
the beta-function equation, $\Lambda$ being the standard convention,
and $\tilde{\Lambda}$ being the convention favoured in
Ref.~\cite{r23}.

Equation~(\ref{eq:CORGIallo}) can easily be adapted to provide a
prediction for the unpolarized Bjorken sum rule of Eq.~(\ref{Ub}).
We simply use the equivalent coefficients of $\MU(a)$ in
Eqs.~(\ref{eq:X2}) and (\ref{eq:X3}) and the following coupling,
\begin{eqnarray}
a_{0}(Q)=\frac{-1}{c\Big{(}1+W_{-1}\Big{[}-\frac{1}{e}\Big{(}\frac{Q^{2}}{\Lambda_{\MU}^{2}}\Big{)}^{-b/2c}\Big{]}\Big{)}},\label{eq:aou}
\end{eqnarray}
where,
\begin{eqnarray}
\Lambda_{\MU}&=&\tilde{\Lambda}_{\rmt\MSbar}\exp(u_{1}^{\rmt\MSbar}/b).
\label{eq:Lu}
\end{eqnarray}

Equation~(\ref{eq:Kinf}) can be modified so that is resums the leading-$b$
components of Eq.~(\ref{eq:CORGIallo}) and essentially becomes an all-orders
CORGI result. We define the coupling $a_{v}$ through,
\begin{eqnarray}
\frac{1}{a_{v}}=\frac{1}{a_{0}}+k_{1}^{(1)}b,\label{eq:av}
\end{eqnarray}
where $k_{1}^{(1)}$ is calculated in the V-scheme. The coupling $a_{0}$ in Eq.~(\ref{eq:av}) is that of
Eq.~(\ref{eq:aok}). The all-orders CORGI result can now be obtained from Eq.~(\ref{eq:Kinf})
using Eq.~(\ref{eq:av}),
\begin{eqnarray}
\tilde{\MK}_{\infty}(a_0)&\equiv&a_{0}+\sum_{n=2}^{\infty}X_{n}^{(n)}b^{n}a_{0}^{n+1}\nn\\
&=&\MK_{\infty}(a_{v}),
\label{eq:KCORinf}
\end{eqnarray}
where $X_{n}^{(n)}$, in analogy with Eq.~(\ref{eq:leadingb}), is the leading-$b$ part of $X_{n}$, for example:
\begin{eqnarray}
X_{2}^{(2)}&=&k_{2}^{(2)}-\left(k_{1}^{(1)}\right)^{2},\\
X^{(3)}_{3}&=&k_{3}^{(3)}-3k_{1}^{(1)}k_{2}^{(2)}+2\left(k_{1}^{(1)}\right)^{3}.
\end{eqnarray}
We can also improve upon $\tilde{\MK}_{\infty}(a_{0})$ by adding to
Eq.~(\ref{eq:KCORinf}) the known
 sub-leading-$b$ part of $X_{2}$,
\begin{eqnarray}
\tilde{\MK}_{\infty+}(a_0)=\tilde{\MK}_{\infty}(a_0)+(X_{2}-X_{2}^{(2)}b^{2})a_{0}^{3}(Q).
\label{eq:ALLO+}
\end{eqnarray}
Equation~(\ref{eq:ALLO+}) now contains all the information we have
about the perturbative coefficients for $\MK(a)$ at all-orders.
Analogous expressions also hold for $\tilde{\MU}_{\infty}(a)$ and
$\tilde{\MU}_{\infty+}(a)$.

In Ref.~\cite{r12}, we noted that the one chain result of
Eq.~(\ref{eq:Kinf}), is finite at the Landau pole $(Q=\Lambda)$, and
remains so for values of $Q$ below $\Lambda$. Remarkably, this
finiteness at $Q=\Lambda$ holds when we use the 't Hooft coupling of
Eq.~(\ref{eq:aok}). Furthermore, both of these results have the same
values at their respective Landau poles, i.e.
\begin{eqnarray}
\MK(a)\Big{|}_{Q=\Lambda}=\MK(a_{0})\Big{|}_{Q=\Lambda_{\MK}}=-\frac{8}{9b}\ln 2.
\label{eq:finite}
\end{eqnarray}
Similar relations apply to $\MU(a)$ and also to the Adler-D
function. This conclusion is altered slightly for the actual CORGI
result of Eq.~(\ref{eq:KCORinf}) because of the use of $a_{v}(Q)$.
In this case the result remains finite at $Q=\Lambda_{\MK}$, but has
a different value to that in Eq.~(\ref{eq:finite})

We can also consider NLO and NNLO CORGI, fixed-order approximations
which can then be compared with all-orders resummations,
\begin{eqnarray}
\tilde{\MK}_{\rmt{NLO}}(a_0)&=&{a}_{0}(Q),
\label{eq:NLO}\\
\tilde{\MK}_{\rmt{NNLO}}(a_0)&=&{a}_{0}(Q)+{X}_{2}{a}_{0}^{3}(Q).
\label{eq:NNLO}
\end{eqnarray}
By comparing these with the all-orders predictions, we can asses the
reliability of fixed-order perturbative predictions. Before doing so
we consider the accuracy of the leading-$b$ approximation for the
CORGI invariants $X_2(\MK)$ and $X_2(\MU)$, as we did for the
perturbative coefficients earlier. For ${X}_{2}(\MK)$ one finds that
\begin{eqnarray}
b{X}_{2}(\MK)&=&-0.0221{N}_{f}^{3}+1.54{N}_{f}^{2}-29.1{N}_{f}+98.6,
\\
b{X}_{2}^{(L)}(\MK)&=&-0.0221{N}_{f}^{3}+1.09{N}_{f}^{2}-18.1{N}_{f}+99.4.
\end{eqnarray}
Notice that the invariants expanded in powers of $b$ contain a
$b^{-1}$ term \cite{r23a} and so it is necessary to multiply by a
factor of $b$ to obtain the expansion in powers of $N_f$. As we
found for the perturbative coefficients in Eqs.~(20-27) there is
reasonably good agreement in sign and magnitude of the coefficients
with astonishingly good $\sim 1\%$ level agreement for the $N_f=0$
large-$N$ coefficient in this case. The numerical values of the
$X_2(\MK)$ invariants for $N_f\{ 0,1,2,3,4,5\}$ are
\begin{eqnarray}
X_2(\MK)&=&\{17.9,13.7,9.61,5.48,1.33,-2.87\},
\\
X_2^{(L)}(\MK)&=&\{ 18.1,15.9,14,12.1,10.4,8.78\}.
\end{eqnarray}
We see good agreement for $N_f=0$ which gradually worsens for
increasing $N_f$. Note that $X_2(\MK)$ changes sign between $N_f=4$
and $N_f=5$, and so big cancellations are involved in this region.
For $X_2(\MU)$ we find
\begin{eqnarray}
b{X}_{2}(\MU)&=&-0.0139{N}_{f}^{3}+1.18{N}_{f}^{2}-25.1{N}_{f}+87.8,
\\
b{X}_{2}^{(L)}(\MU)&=&-0.0139{N}_{f}^{3}+0.688{N}_{f}^{2}-11.3N_f+62.4.
\end{eqnarray}
The agreement appears to be significantly worse in this case although the pattern
of signs is reproduced. The numerical values are
\begin{eqnarray}
X_2(\MU)&=&\{16,12.3,8.7,5.01,1.22,-2.69\},
\\
X_2^{(L)}(\MU)&=&\{11.3,10,8.76,7.59,6.51,5.51\}.
\end{eqnarray}

\section{Comparison of fixed-order and all-orders predictions}\vspace{-11pt}
The data available for $K_{GLS}(Q^{2})$ and $K_{uBj}(Q^{2})$ span a
range of energy which includes the bottom quark mass threshold. It
is therefore necessary for us to choose a method for evolving the
above expressions through this threshold. We adopt the approach
detailed in \cite{r24}. At a particular energy scale we treat all
quarks with masses less than that scale as `active' but massless and
we ignore quarks with masses greater than that scale. As a
consequence, the coefficients $k_{n}$ and $u_{n}$ are now
$N_{f}$-dependent and hence they will depend on the c.m. energy
scale, $Q$.

We also perform matching of the coupling at $Q^{2}=m_{b}^{2}$. At LO
and NLO this amounts to demanding continuity of the coupling at the
threshold but at NNLO and beyond, this continuity is violated. This
matching forces us to adopt different values of the scale parameter
in different $N_{f}$ regions. This is governed by the following
equations,
\begin{eqnarray}
\Lambda_{N_{f}+1}^{2}&=&\Lambda_{N_{f}}^{2}\left(\frac{m_{N_{f}+1}^{2}}{\Lambda_{N_{f}}^{2}}\right)^{1-\frac{b^{N_{f}}}{b^{N_{f}+1}}}\times\exp\left(\frac{\delta_{\rmt{NLO}}+\delta_{\rmt{NNLO}}}{2b^{N_{f}+1}}\right),
\label{eq:matching}
\end{eqnarray}
where $\delta_{\rmt{NLO}}$ and $\delta_{\rmt{NNLO}}$ can be obtained
from the results in Ref.~\cite{r24}. Here, $\Lambda_{N_{f}}$ is the
scale parameter in the region where $N_{f}$ quarks are active,
$m_{N_{f}}$ is the pole mass of the $f$ quark and $b^{N_{f}}$ is
simply $b$ evaluated for $N_{f}$ quark flavours.

We must be careful how we apply this matching to the different
results we have obtained. Equation (\ref{eq:matching}) is an
$\MSbar$ result and hence we carry out the matching for $\LMS$ and
then convert to the various other scales we have defined in
Eqs.~(\ref{eq:Lv}), (\ref{eq:Lk}) and (\ref{eq:Lu}).
$\tilde{\MK}_{\rmt{NLO}}(a)$ is an NLO result and hence, when
applying Eq.~(\ref{eq:matching}) to this result, we omit
$\delta_{\rmt{NNLO}}$. The results in Eqs.~(\ref{eq:KCORinf}),
(\ref{eq:ALLO+}) and (\ref{eq:NNLO}) are all, at least, NNLO and
hence we use the full result from Eq.~(\ref{eq:matching}).

\begin{figure}
\begin{center}
\includegraphics[angle=270,width=.7\textwidth]{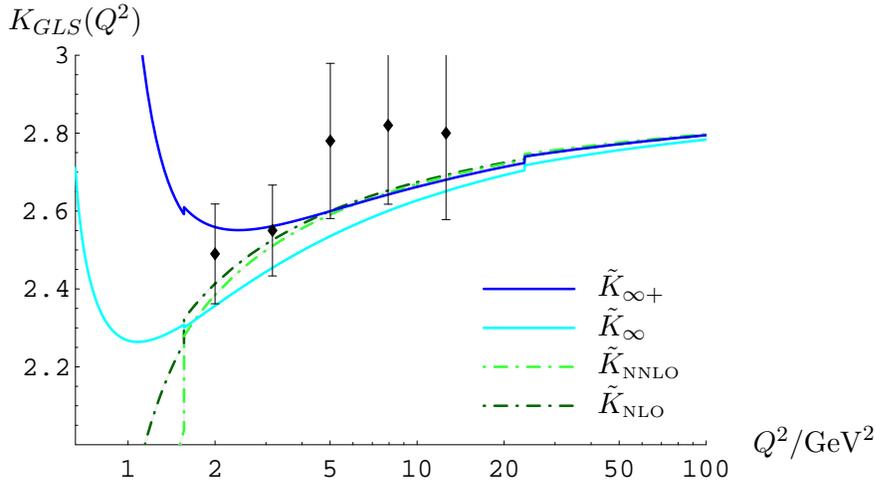}%KGLS2.ps
\caption{Various predictions for the GLS sum rule,
  superimposed on to experimental data.}
\label{fig1}
\end{center}
\end{figure}
\begin{figure}
\begin{center}
\includegraphics[angle=270,width=.7\textwidth]{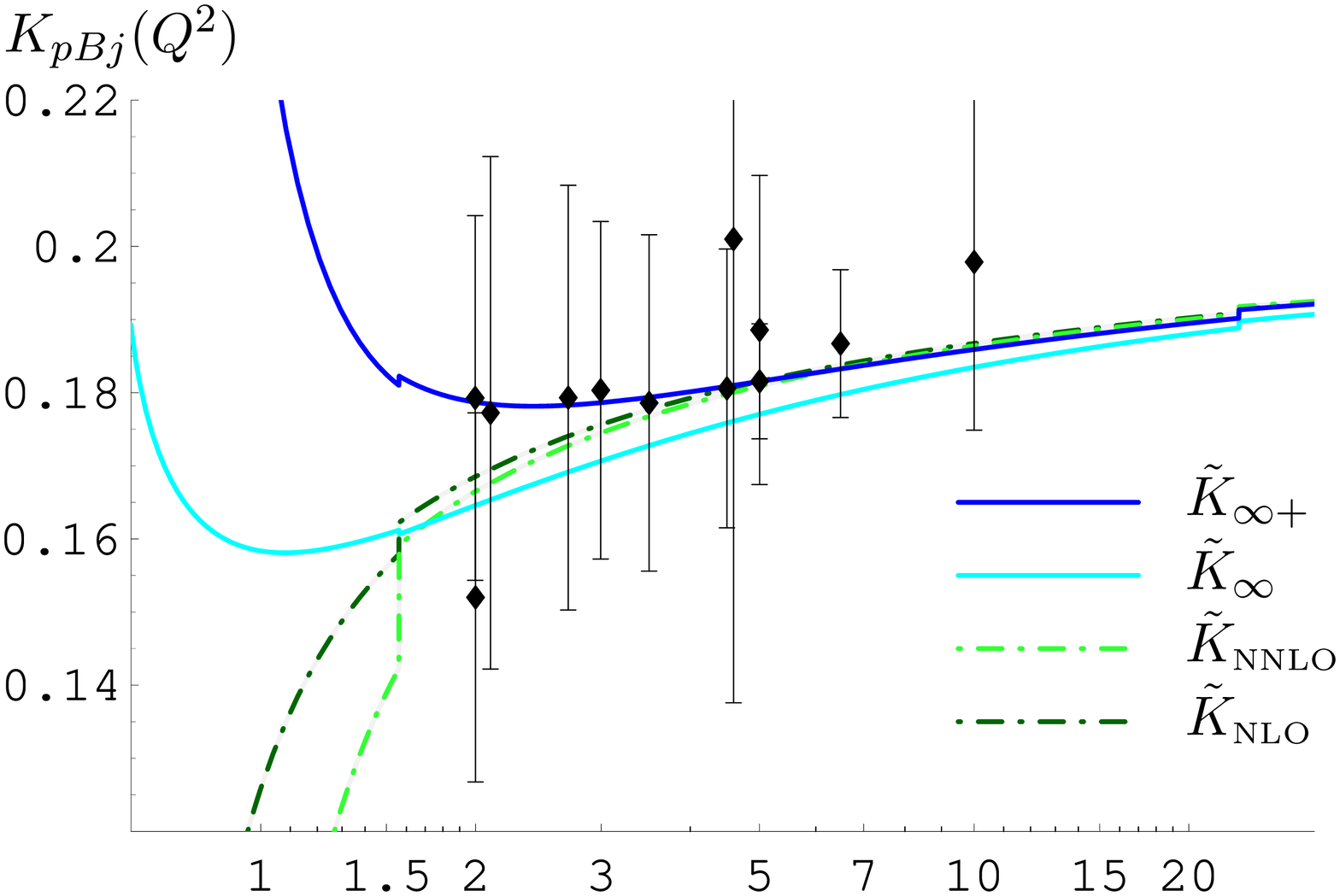}%Kbj2.ps
\caption{Various predictions for the polarized Bjorken sum rule,
  superimposed on to experimental data.}
\label{fig2}
\end{center}
\end{figure}
\begin{figure}
\begin{center}
\includegraphics[angle=270,width=.7\textwidth]{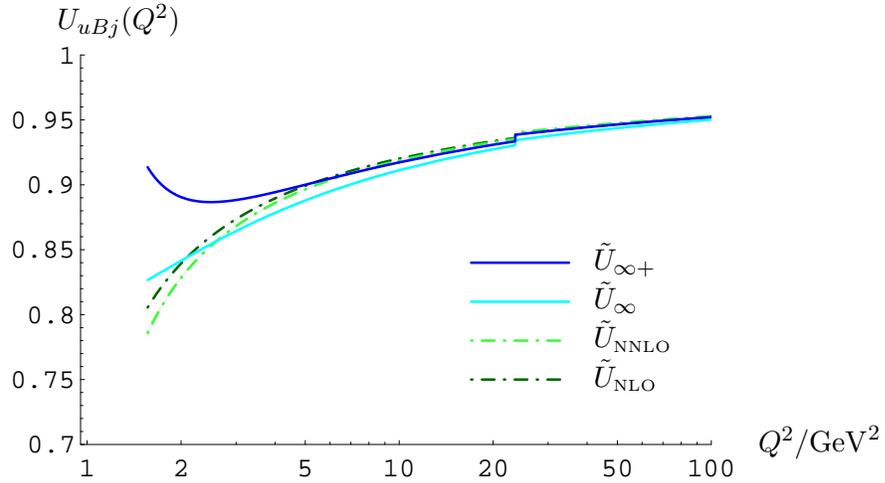}%Upert.ps
\caption{Various predictions for the unpolarized Bjorken sum rule.}
\end{center}
\label{f:fig3}
\end{figure}
We wish to compare these results with relevant experimental data.
Data is available for the GLS sum rule \cite{r8} and the polarized
Bjorken sum rule \cite{r9,r10,r11,r11a}, where the points we plot
are those arising from the analysis of \cite{r11b}. We plot the data
in Figs.~\ref{fig1} and \ref{fig2} along with our predictions for
the full observables. No experiment has so far measured the
unpolarized Bjorken sum rule, however the possibility exists that it
may be extracted from experiments at a future neutrino factory
\cite{r25}. We have taken ${\Lambda}^{(5)}_{\rmt\MSbar}=207$ MeV,
corresponding to the world average value ${\alpha}_{s}(M_Z)=0.1176$
\cite{r15a}.

Within the considerable error bars we see that the different
versions of CORGI all fit the data tolerably well. The important
conclusion is that below $Q^2\sim{5}$ ${\rm{GeV}}^2$ the different
approaches give drastically different predictions. This indicates
that fixed-order perturbation theory cannot be trusted for these
lower energies. Even though we have no experimental data points for
${U}_{uBj}(Q^2)$ we have plotted the different versions of CORGI in
Fig.~\ref{f:fig3}. We see that fixed-order perturbation theory
cannot be trusted below $Q^2\sim{2}$ ${\rm{GeV}}^2$. A surprising
feature evident from Figs.~1 - 3 is that even at
$Q^2=2{\rm{GeV}}^{2}$ the fixed-order NLO and NNLO results remain
extremely close, leading to the superficial conclusion that
fixed-order perturbation theory is working well. In fact as can be
seen from Eqs.~(54,56) the CORGI invariant $X_2$ for both $\MK$ and
$\MU$ changes sign between $N_f=4$ and $N_f=5$ and so fortuitously
is very small in this region. When $X_3^{(L)}$ and higher invariants
are resummed the potential unreliability of fixed-order perturbation
theory is revealed.

We can contrast these CORGI resummations with a leading-$b$
resummation in the \MSbar~scheme. If we choose the scale $\mu=xQ$ we
have, analogous to Eq.~(\ref{eq:KCORinf}),
\begin{eqnarray}
{\MK}_{\infty}^{\rmt\MSbar}(x)&\equiv&a(xQ)+\sum_{n=1}^{\infty}k_{n}^{(n)}(x)b^n{a}^{n+1}(xQ)
\\
&=&{\MK}_{\infty}({a}_{v}(x))\;.
\end{eqnarray}
Here $a(xQ)$ denotes the full higher-loop (three-loop if matching to
NNLO) \MSbar~coupling at scale $\mu=xQ$, and the $k_n(x)$
coefficients are the \MSbar~coefficients with scale $\mu=xQ$. The
coupling ${a}_{v}(x)$ is then defined by,
\begin{equation}
\frac{1}{{a}_{v}(x)}=\frac{1}{a(xQ)}-b\left(\ln x+\frac{5}{6}\right)\;.
\end{equation}
We can match this all-orders resummation to the exact NLO and NNLO
perturbative coefficients, obtaining, in
analogy with Eq.~(\ref{eq:ALLO+}),
\begin{equation}
{\MK}^{\rmt\MSbar}_{\infty
  +}(x)={\MK}^{\rmt\MSbar}_{\infty}(x)+({k}_{1}(x)-b{k}_{1}^{(1)}(x)){a}^{2}(xQ)+({k}_{2}(x)-b^2{k}_{2}^{(2)}(x)){a}^{3}(xQ),
\label{eq:MSleadingb}
\end{equation}
where in this $a(Q)$ is the approximated three-loop coupling, given in \cite{r15a}.

\begin{figure}
\begin{center}
\includegraphics[angle=270,width=.7\textwidth]{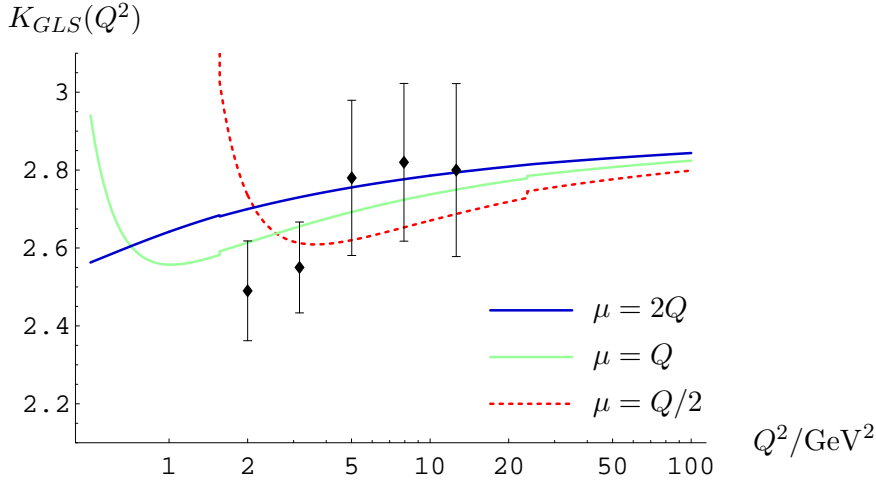}%GLSmatch.ps
\caption{All-orders leading-$b$ prediction for the GLS sum rule.
Here we use
  the \MSbar~result of Eq.~(\ref{eq:MSleadingb}) with renormalization scales of $\mu=2Q,Q$ and $Q/2$.}
\label{f:fig4}
\end{center}
\end{figure}

The resulting prediction is plotted in Fig.~\ref{f:fig4} for the GLS
sum rule. Three different matching scales corresponding to
$x=2,1,\frac{1}{2}$ were chosen. As can be seen, the  matched
resummed perturbative result is hopelessly $x$-dependent. The CORGI
result $\tilde{\MK}_{\infty +}(a)$ corresponds to the choice
$x={e}^{-{k}_{1}(1)/b}$. However, we note that no NLO matching is
required in the CORGI approach, since an infinite set of
RG-predictable terms involving $k_1$ have been resummed to
all-orders, yielding a $\mu$-independent result.

\section{Non-perturbative corrections}\vspace{-11pt}
In addition to perturbative corrections it is expected that there
will be non-perturbative, Higher-Twist (HT) corrections to the sum
rules. In a recent paper \cite{r12} we have studied the infra-red
freezing of Euclidean observables, including the sum rules, in the
language of the one-chain term of the skeleton expansion. The
leading-$b$ resummations we have discussed can be written in the
form,
\begin{equation}
{\MK}_\infty(Q^2)=\int_{0}^{1}{dt}\;{\omega}^{\rmt{IR}}_{\MK}(t)a(t{Q}^{2})+\int_{1}^{\infty}{dt}\;{\omega}^{\rmt{UV}}_{\MK}(t)a(t{Q}^{2})\;.
\end{equation}
Here $\omega_{\MK}(t)$ is the characteristic function which is
piecewise continuous at $t=1$. The two terms respectively reproduce
the IR renormalon and UV renormalon contributions in the Borel sum,
but the skeleton expansion result is also defined for
$Q^2<\Lambda^2$ where the standard Borel representation breaks down.
Remarkably, there is continuity and finiteness at $Q^2=\Lambda^2$.
By considering the compensation of ambiguities between perturbative
and non-perturbative corrections alluded to above we are led to an
expression for HT corrections in terms of the characteristic
function,
\begin{equation}
{\MK}_{\rmt{HT}}(Q^2)=\kappa\frac{{\Lambda}^{2}}{Q^2}\omega^{\rmt{IR}}_{\MK}\left(\frac{\Lambda^{2}}{Q^2}\right)\;.
\label{eq:kHT}
\end{equation}
Here $\kappa$ is a single overall non-perturbative constant. This
result is obtained from the V-scheme result of Eq.~(\ref{eq:Kinf})
and hence the appropriate scale parameter for Eq.~(\ref{eq:kHT}) is
$\Lambda_{{\rm V}}$. A similar expression holds for
${\MU}_{\rmt{HT}}(Q^{2})$ in terms of
${\omega}^{\rmt{IR}}_{\MU}(t)$.

The characteristic functions $\omega_{\MK}(t)$ and $\omega_{\MU}(t)$
can be determined from their Borel transforms,  using a result from
\cite{r12}. $\omega_{\MK}(t)$ has the form,
\begin{eqnarray}
\omega_{\MK}^{\rmt{IR}}(t)&=&\frac{8}{9}-\frac{5}{9}t,\\
\omega_{\MK}^{\rmt{UV}}(t)&=&\frac{4}{9t^{2}}-\frac{1}{9t^{3}},
\end{eqnarray}
and the equivalent expression for $\MU$ is,
\begin{eqnarray}
\omega_{\MU}^{\rmt{IR}}(t)&=&\frac{4}{3}-t,\\
\omega_{\MU}^{\rmt{UV}}(t)&=&\frac{1}{3t^{3}}.
\end{eqnarray}

\begin{figure}
\begin{center}
\includegraphics[angle=270,width=.7\textwidth]{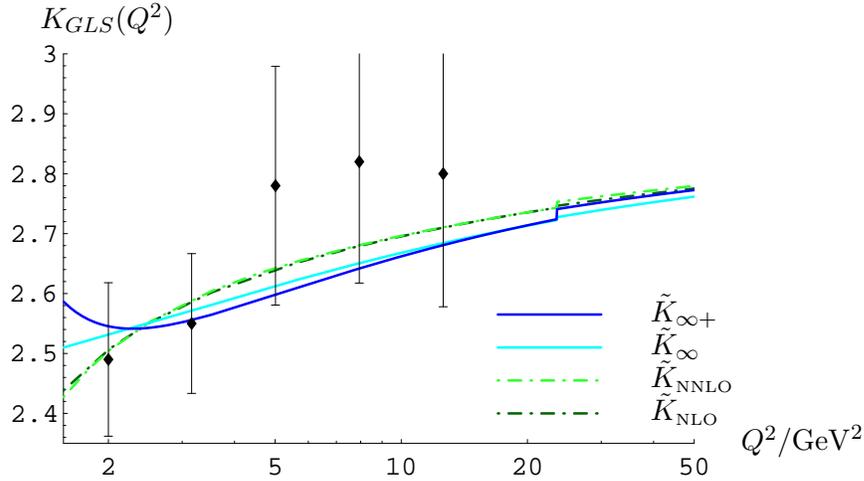}%GLSfit2.ps
\caption{Predictions for the GLS sum rule, including both perturbative and
  non-perturbative corrections, fitted to the data by varying the parameter $\kappa_{GLS}$.}
\label{f:GLSfit}
\end{center}
\end{figure}

\begin{figure}
\begin{center}
\includegraphics[angle=270,width=.7\textwidth]{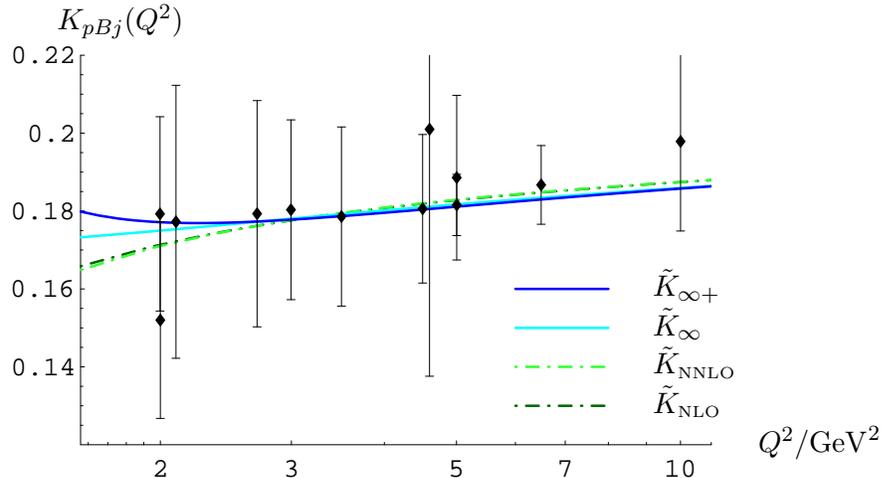}%pBjfit2.ps
\caption{Predictions for the polarized Bjorken sum rule, including both perturbative and
  non-perturbative corrections, fitted to the data by varying the parameter $\kappa_{pBj}$.}
\label{f:pBjfit}
\end{center}
\end{figure}

In Figs.~\ref{f:GLSfit} and \ref{f:pBjfit} we take the four
expressions in Eqs.~(\ref{eq:KCORinf}), (\ref{eq:ALLO+}),
(\ref{eq:NLO}) and (\ref{eq:NNLO}), supplemented by the
non-perturbative term in Eq.~(\ref{eq:kHT}), and perform fitting to
experimental data for both the GLS and the polarized Bjorken sum
rules separately. We fix ${\Lambda}^{(5)}_{\rmt\MSbar}=207$ MeV, as
before, and use $\chi^{2}$ fitting to obtain the optimal value of
the non-perturbative parameters in each case, $\kappa_{GLS}$ and
$\kappa_{pBj}$. The fitted parameters are summarized in Table 1. We
note that the coefficient of a $1/Q^{2}$ (twist-4) power correction
to either Eq.~(\ref{Kg}) or (\ref{Kp}) corresponds to a value of
$-8\Lambda_{V}^{2}\kappa_{GLS}/3$ or
$-\frac{4}{27}|g_{A}/g_{V}|\Lambda_{V}^{2}\kappa_{pBj}$,
respectively. The corresponding values in ${\rm{GeV}}^{2}$ are
presented in Table 1. For comparison the central values resulting
from a three-point function QCD sum rules fit \cite{r25a,r25b} are
$-0.294$ and $-0.013$, respectively. We see that the power
corrections in Table 1 resulting from fitting to the resummed
${\tilde{\MK}}_{\infty +}(a)$ all-orders resummations are
significantly smaller, although the errors are large. Connections
between power corrections for the three DIS sum rules have also been
explored in \cite{r26}.
\begin{table}
\begin{center}
\begin{tabular}{|c|c|c|c|c|c|c|}
 \hline&&&&&&\\[-11pt]
 &$\kappa_{GLS}$&$-\frac{8}{3}\kappa_{GLS}\Lambda^{2}_{V}$&$\frac{\chi_{GLS}^{2}}{\rm d.o.f.}$&$\kappa_{pBj}$&$-\frac{4}{27}\left|\frac{g_{A}}{g_{V}}\right|\kappa_{pBj}\Lambda^{2}_{V}$&$\frac{\chi_{pBj}^{2}}{\rm d.o.f.}$
 \\[3pt]\hline
$\tMK_{\rmt{NLO}}$ & -0.166      &$0.1007  \pm 0.12$ &$1.256/4$ & -0.06255 &$0.002647\pm 0.011$ & 1.173/11 \\
$\tMK_{\rmt{NNLO}}$ & -0.217     &$0.1313  \pm 0.12$ &$1.238/4$ & -0.1085 &$ 0.004593\pm 0.011$ &  1.162/11 \\
$\tMK_{\infty}$ & -0.33    &$0.1997  \pm 0.12$ &$1.804/4$ & -0.2823  &$0.01195\pm  0.011$ & 1.456/11 \\
$\tMK_{\infty+}$  & 0.0216 &$-0.01311\pm 0.12$ &$2.089/4$ & 0.03957  &$-0.001674\pm 0.011$ & 1.662/11 \\\hline
\end{tabular}
\end{center}
\caption{Fitted values of the non-perturbative constants $\kappa_{GLS}$ and $\kappa_{pBj}$, together with their
 respective $\chi^{2}$ per degree of freedom.}
\label{tab1}
\end{table}

\section{Conclusions}\vspace{-11pt}
By comparing leading-$b$ resummations based on the CORGI approach we
attempted to infer the validity of fixed-order perturbation theory
for the GLS and pBj DIS sum rules at small energy scales
$Q^2\sim{3}\;{\rm{GeV}}^{2}$, where data has previously been used to
extract ${\alpha}_{s}(M_Z)$  \cite{r15a}. Figures 1 and 2 indicate
that fixed-order perturbation theory and resummed all-orders
predictions start to differ drastically for energies below
$Q^2\sim{5}\;{\rm{GeV}}^{2}$. The use of the CORGI approach ensures
that all RG-predictable scale-dependent logarithms are resummed to
all-orders, thus avoiding the need for NLO matching which would
otherwise make the conclusions about the validity of fixed-order
perturbation theory dependent on the chosen matching scale (as
illustrated in Fig.~4).

We performed fits to data for power corrections using the model
proposed in Ref.~\cite{r12} based on the skeleton expansion
characteristic function. The size of power corrections inferred by
fitting the all-orders CORGI resummations was much smaller than the
central values obtained from three-point QCD sum rule estimates of
Refs.~\cite{r25a,r25b}.

We should note that in a recent interesting paper \cite{r6h} a
renormalon analysis of pBj and GLS sum rules was also carried out.
The aim in that case was to perform a sophisticated matching of
perturbative and non-perturbative effects in the renormalon
subtracted approach.

\section*{Acknowledgements}\vspace{-11pt} We would like to thank Andrei Kataev for useful discussions on the topic
of DIS sum rules and power corrections. P.M.B. gratefully
acknowledges the receipt of a PPARC UK studentship.


\begin{thebibliography}{99}
\setlength{\baselineskip}{16.5pt}
\setlength{\parskip}{0pt}


\bibitem{r1} For a review see: M. Beneke, Phys. Rep. {\bf 317}, 1 (1999), [hep-ph/9807443];
M. Beneke and V. M. Braun, published in `The Boris Ioffe
Festschrift- At the Frontier of Particle Physics/Handbook of QCD'',
edited by M. Shifman (World Scientific, Singapore, 2001),
[hep-ph/0010208]

\bibitem{r2} C. N. Lovett-Turner and C. J. Maxwell, Nucl. Phys. \textbf{B452} (1995) 188, [hep-ph/9505224]

\bibitem{r3} D. J. Broadhurst and A. G. Grozin, Phys. Rev. {\bf D52} (1995)
  4082,  [hep-ph/9410240].

\bibitem{r4} M. Beneke and V. M. Braun , Phys. Lett. {\bf B348}
  (1995) 513, [hep-ph/9411229].

\bibitem{r5} M. Neubert, Phys. Rev. {\bf D51} (1995) 5924, [hep-ph/9412265].

\bibitem{r5a} J. Chyla, A. Kataev and S. Larin Phys. Lett. {\bf B267} (1991) 269.

\bibitem{r6a} D. G. Tonge and C. J. Maxwell, Nucl. Phys. {\bf B481} (1996) 681,
  [hep-ph/9606392];

D. G. Tonge and C. J. Maxwell, Nucl. Phys. {\bf B535} (1998) 19,
[hep-ph/9705314].

\bibitem{r6A} D. M. Howe and C. J. Maxwell, Phys. Rev. {\bf D70} (2004) 014002. [hep-ph/0303163].

\bibitem{r6b} J. Chyla and A.L. Kataev, Phys. Lett. {\bf B297} (1992) 385. [hep-ph/9209213].

\bibitem{r6c} J. Ellis, E. Gardi, M. Karliner and M.A. Samuel, Phys. Lett. {\bf B366} (1996) 268. [hep-ph/9609312].

\bibitem{r6d} G. Altarelli, R. D. Ball, S. Forte and G. Ridolfi, Nucl. Phys. {\bf B496} (1997) 337. [hep-ph/9701289].

\bibitem{r6e} T. Lee Phys. Rev.{\bf D56} (1997) 1091. [hep-ph/9611010];

T. Lee Phys. Lett. {\bf B462} (1999) 1. [hep-ph/9908225]

\bibitem{r6f} K. A. Milton, I. L. Solovtsov and O. P. Solovtsova, Phys. Rev. {\bf D60} (1999) 016001.

\bibitem{r6g} C. Contreras, G. Cvetic, K. S. Jeong and T. Lee, Phys. Rev. {\bf D66}, 054006 (2002). [hep-ph/0203201]

\bibitem{r6h} F. Campanario and A. Pineda, Phys. Rev. {\bf D72} (2005) 056008. [hep-ph/0508217].

\bibitem{r6} G. Grunberg, Phys. Lett. {\bf B95} (1980) 70;

G. Grunberg, Phys. Rev. {\bf D29} (1984) 2315.

\bibitem{r7} C. J. Maxwell [hep-ph/9908463];

C. J. Maxwell and A. Mirjalili, Nucl. Phys. {\bf B577} (2000) 209.
[hep-ph/0002204].

\bibitem{r7a} C. J. Maxwell and A. Mirjalili, Nucl. Phys. {\bf B611} (2001) 423, [hep-ph/0103164].

\bibitem{r8} J. H. Kim and D. A. Harris et al. Phys. Rev. Lett. \textbf{81}
  (1998) 3595, [hep-ex/9808015].

\bibitem{r9}  K. Abe et al.[E143 collaboration], Phys. Rev.{\bf D58}, (1998) 112003. [hep-ph/9802357].

\bibitem{r10} B. Adeva, et al. [Spin Muon Collaboration] Phys. Rev \textbf{D58} (1998) 112002.

\bibitem{r11} P. L. Anthony et al. [E155 Collaboration], Phys. Lett.{\bf B493} (2000) 19. {hep-ph/0007248]

\bibitem{r11a} A. Airapetian et al. [HERMES collaboration], Eur. Phys. J. {\bf C26} (2003) 527. [hep-ex/0210047]

%\cite{Deur:2004ti}
\bibitem{r11b}
  A.~Deur {\it et al.},
  %``Experimental determination of the evolution of the Bjorken integral at  low
  %Q**2,''
  Phys.\ Rev.\ Lett.\  {\bf 93} (2004) 212001.
  [hep-ex/0407007].
  %%CITATION = HEP-EX 0407007;%%

\bibitem{r12} P.~M.~Brooks and C.~J.~Maxwell,
  %``Infrared freezing of Euclidean QCD observables,''
  Phys.\ Rev.\  D {\bf 74} (2006) 065012
  [hep-ph/0604267].

\bibitem{r13}  D.~J.~Gross and C. H. Llewellyn Smith, Nucl. Phys. \textbf{B14} (1969) 337.

\bibitem{r14}   J.~D.~Bjorken, Phys. Rev. \textbf{148} (1966) 1467; \textbf{D1} (1970) 1376.

\bibitem{r15}   J. D. Bjorken, Phys. Rev. \textbf{179} (1969) 1547.

\bibitem{r15a}
  S.~Eidelman {\it et al.}  [Particle Data Group],
  %``Review of particle physics,''
  Phys.\ Lett.\ B {\bf 592} (2004) 1.
  %%CITATION = PHLTA,B592,1;%%

\bibitem{r15b} S. G. Gorishny and S. A. Larin, Phys. Lett. {\bf B172} (1986) 109;

S.A. Larin and J. A. M. Vermaseren, Phys. Lett. {\bf B259} (1991)
345.

\bibitem{r15c} K. G. Chetyrkin, S. G. Gorishny, S. A. Larin and F. V. Tkachov,Phys. Lett. {\bf B137} (1984) 230;

S.A. Larin, F. V. Tkachov, and J. A. M. Vermaseren, Phys. Rev. Lett.
{\bf 66} (1991) 862.

\bibitem{r15d} C.J. Maxwell, Phys. Lett. {\bf B409} (1997) 382. [hep-ph/9706231].

\bibitem{r15e} M. Beneke, V. Braun and N. Kivel, Phys. Lett. {\bf B404} (1997) 315. [hep-ph/9703389].

\bibitem{r16}  D. J. Broadhurst and A. L. Kataev, Phys. Lett. \textbf{B315}
  (1993) 179, [hep-ph/9308274].


\bibitem{r17}  D. J. Broadhurst and A.L. Kataev, Phys. Lett. \textbf{B544} (2002) 154, [hep-ph/0207267].

\bibitem{r19}  E. Gardi, G. Grunberg and M. Karliner, JHEP {\bf 9807} (1998)
  007, [hep-ph/9806462].

\bibitem{r20}  M. A. Magradze, Int. J. Mod. Phys. {\bf A15},  (2000) 2715, [hep-ph/9911456].

\bibitem{r21}  G 't Hooft, in Deeper Pathways in High Energy Physics, proceedings of
Orbis Scientiae, 1977, Coral Gables, Florida, edited by A. Perlmutter and L.F. Scott
(Plenum, New York, 1977).

\bibitem{r22}  R. M.~Corless,~G. H.~Gonnet,~D. E. G~Hare,~D. J.~Jeffrey and D. E. Knuth, ``On~the
~Lambert~$W$~function'', Advances~in~Computational~Mathematics~{\bf 5}~(1996)~329,
available from {\tt http://www.apmaths.uwo.ca/$\sim$djeffrey/offprints.html.}

\bibitem{r23}  P. M. Stevenson, Phys. Rev. {\bf D23}, (1981) 2916.

\bibitem{r23a} G. Grunberg, Phys. Rev. {\bf D46} (1992) 2228.

\bibitem{r24}  W. Bernreuther and W Wetzel, Nucl. Phys. \textbf{B197} (1982) 228; Erratum {\it ibid.} {\bf B513} (1998) 758.

\bibitem{r25}  S. L. Alekhin and A. L. Kataev, J. Phys. {\bf G29} (2003) 1993, [hep-ph/0209165].

\bibitem{r25a} V. M. Braun and A. V. Kolesnichenko, Nucl. Phys. {\bf B283} (1987) 723.

\bibitem{r25b} I. L. Balitsky, V. M. Braun and A. V. Kolesnichenko, Phys. Lett.{\bf B242} (1990) 245;
Erratum {\it ibid.} {\bf B318} (1993) 648.

\bibitem{r26} A. L. Kataev, Mod. Phys. Lett. {\bf A20} (2005)} 2007. [hep-ph/0505230].
\end{thebibliography}
\end{document}